\documentclass[sigconf]{acmart}

\usepackage{lipsum}
\usepackage{graphicx}
\usepackage{hyperref}
\usepackage[font=small, skip=0pt]{caption}

\usepackage{subcaption}

\begin{document}

\title{Predicting the Impact of Batch Refactoring Code Smells on Application Resource Consumption\\
}

\author[1]{Asif Imran}
\author[2]{Tevfik Kosar}
\author[3]{Jaroslaw Zola}
\author[4]{Fatih Bulut}
\begin{abstract}
\textbf{Background:} Automated batch refactoring has become a de-facto mechanism to restructure software that may have significant design flaws negatively impacting the code quality and maintainability. Although automated batch refactoring techniques are known to significantly improve overall software quality and maintainability, their impact on resource utilization is not well studied. \\
\textbf{Aims:} This paper aims to bridge the gap between batch refactoring code smells and consumption of resources. It determines the relationship between software code smell batch refactoring, and resource consumption. Next, it aims to design algorithms to predict the impact of code smell refactoring on resource consumption.\\
\textbf{Method:} This paper investigates 16 code smell types and their joint effect on resource utilization for 31 open source applications. It provides a detailed empirical analysis of the change in application CPU and memory utilization after refactoring specific code smells in isolation and in batches. This analysis is then used to train regression algorithms to predict the impact of batch refactoring on CPU and memory utilization before making any refactoring decisions. \\
\textbf{Results:} Experimental results also show that our ANN-based regression model provides highly accurate predictions for the impact of batch refactoring on resource consumption. It allows the software developers to intelligently decide which code smells they should refactor jointly to achieve high code quality and maintainability without increasing the application resource utilization. \\
\textbf{Conclusion:} This paper responds to the important and urgent need of software engineers across a broad range of software applications, who are looking to refactor code smells and at the same time improve resource consumption. Finally, it brings forward the concept of resource aware code smell refactoring to the most crucial software applications.

\end{abstract}
\maketitle

\keywords{code smells, software batch refactoring, software resource consumption.}

\section{Introduction}

Modern software development practices suffer from increased pressure to deliver new features in a shorter time to meet the deadlines and compete with peers. Collaborative codebases with a large number of contributors who may have different levels of expertise and coding standards and continuously evolving software without a proper design add to the problem. These practices generally result in code smells, a software behavior that indicates a violation of fundamental design principles and negatively impacts the code's readability, maintainability, and scalability \cite{Fowler}.
%
%
Certain types of code smells may also drain system resources like CPU and memory, resulting in wastage of critical resources, increasing the cost of operating the software, and even degrading the performance of the applications in some cases \cite{perez2014analyzing}. For example, the cyclic dependency code smell violates the acyclic properties of code and introduces loops where it may not be necessary. The enhanced loops will cause repetition of a process in the execution flow and result in excess resource consumption.   


Fixing the code smells is known to improve code quality and maintainability, but it does not always result in better application resource utilization. Some of the code refactoring techniques and tools used during this process can introduce other anomalies that can increase the application's CPU and memory utilization \cite{holzer2009towards}. The majority of the existing work on automated code smell refactoring focuses on correctness \cite{paixao2020behind}, maintainability \cite{vakilian2012use}, and scalability \cite{murphy2008scalable}. Previous research studying the impact of code smell refactoring on resource consumption is quite limited. Also, in modern software coding, smells are refactored in groups, and the impact of batch refactoring code smells on resource consumption requires further exploration \cite{fernandes2019stuck}. Verdecchia et al. \cite{verdecchia2018empirical} performed an exploratory analysis of the impact of code smell refactoring on energy consumption and performance in software applications. They selected five different code smells (feature envy, type checking, long method, god class, and duplicated code), which were automatically detected and refactored in three open-source Java software applications.
Other efforts were limited to the isolated impact of a small segment of code smells, which did not consider the combined impact of refactoring a large number of smells \cite{wang2014platform}. At the same time, previous research studies considered only a handful of applications to analyze the impact \cite{anwar2019evaluating}. 

%

This paper fills a void in this area by providing a comprehensive analysis on the impact of batch refactoring 16 different code smell types on the resource consumption of 31 real-life Java and Python applications. We find that batch refactoring of code smells has a significant impact on both CPU and memory usage. 
%
Depending on the goal of the application developers, this study enables intelligent selection of which smells should be refactored together and which ones not be refactored. If the primary goal is easy to maintain code, then all smells can be refactored. In that case, this study can provide the developers with an estimation of the expected change in resource utilization after batch refactoring. If the concern is not only easy maintenance but also resource consumption of the application, this study will help the developers to intelligently decide which smells to refactor jointly to minimize the resource consumption.


In this study, we used 3 different automated refactoring tools, \textit{Jdeodorant} \cite{fokaefs2011jdeodorant} and \textit{JSparrow} \cite{jsparrow} for Java and \textit{pycharm} for Python \cite{gulabovska2019survey} applications to detect and refactor the code smells. We establish a benchmark where individual types of code smells are detected and refactored in each software, followed by an analysis of CPU and memory consumption impact. Afterward, we conduct a batch refactoring of smells and analyze their collective impact on resource usage. Next, we use the benchmark data to predict the impact of batch smell refactoring on CPU and memory usage. We apply five regression models, namely linear regression, polynomial regression, lasso regression, random forest, and ANN regression, and calculate their accuracy using the mean square error (MSE) and root mean squared error (RMSE) values. Experimental results show that ANN regression outperforms the other models in terms of prediction accuracy. 

The major contributions of this paper include the following:

\begin{itemize}
    \item A detailed impact analysis of refactoring 16 different code smell types on the resource consumption of different Java and Python applications.
    \item An empirical evaluation of the change in resource utilization after auto-refactoring specific code smells in isolation as well as batch refactoring.
    \item A  set  of guiding  principles to select the code smells which will improve resource usage when refactored collectively.
    \item A mechanism based on regression analysis to predict the impact of batch refactoring code smells on CPU and memory utilization before making any refactoring decisions.
     
\end{itemize}


The rest of the paper is organized as follows: Section \ref{methodology} explains the code smell types, the selected applications, workloads and the automated refactoring tools used for this study. 
Section \ref{results} presents the results of the experiments and a summary of our findings, including the impact of batch refactoring and regression-based predictive modeling. Section \ref{relatedwork} discusses the related work in this area, and Section \ref{conclusion} concludes the paper.

\section{Methodology and Experimental Setup}
\label{methodology}


%
Our analysis includes 16 different code smell types, and to the best of our knowledge, this is the most comprehensive study in this area so far. All selected smells can be detected and refactored using off-the-shelf automated refactoring tools. Table \ref{tab00} summarizes the 16 code smells, including their properties, refactoring techniques, and their impact on application resource utilization.

\hyphenpenalty 10000
\begin{table*}
\footnotesize
\begin{tabular}{p{1.1cm}p{3.5cm}p{4.0cm}p{6.5cm}} \hline
\textbf{{Smell Type}} & \textbf{\parbox{3.5cm}{Property}} & \textbf{{Refactoring technique}} & \textbf{{Impact on Resource Utilization}}\\
    \hline
    cyclic dependency&\parbox{3.5cm}{Violates acyclic properties and results in misplaced elements \cite{sarkar2012measuring} }&\parbox{4cm}{Encapsulating all packages in a cycle and assign to single team}&\parbox{6.5cm}{Refactoring prevents the enhanced loops from repeating, thus prevents resource wastage}
    \\
    \hline
    god method&\parbox{3.5cm}{Many activities in a single method \cite{Fowler}} &\parbox{4cm}{Divide the god method into multiple smaller methods} &\parbox{6.5cm}{Multiple processes in a single method cause less inter-method communication, hence preserves resource usage}\\
    \hline
    spaghetti code&\parbox{3.5cm}{Addition of new code without removing obsolete ones \cite{abbes2011empirical}}&\parbox{4cm}{Replace procedural code segments with object oriented design}&\parbox{6.5cm}{Unrefactored code contains length() and size() can have a time complexity of O(n), refactoring results in using \textit{isEmpty()} instead of \textit{length()} and \textit{size()} which has a complexity of \textit{O(1)}}\\
    \hline
    shotgun surgery &\parbox{3.5cm}{Single behavior defined across multiple classes\cite{Fowler}}&\parbox{4cm}{Use Move Method and Move Field to move repetitive class behaviors into a single class}&\parbox{6.5cm}{Refactoring removes the resource-consuming code blocks which were applied in multiple locations}\\
    \hline
    god class &\parbox{3.5cm}{One class aims to do activities of many classes \cite{Fowler}}&\parbox{4cm}{Divide the large class into smaller classes}&\parbox{6.5cm}{Refactoring causes greater inter-class communication, thus increasing resource consumption}\\
    \hline
    lazy class &\parbox{3.5cm}{The class does not do enough activity and can be easily replaced \cite{Fowler}}&\parbox{4cm}{Use diamond operators to remove re-implementation of interface}&\parbox{6.5cm}{Refactoring lazy class prevents the consumption of excess resource due to context switching from this class to the other classes}\\
    \hline
    refused bequest &\parbox{3.5cm}{When the child classes of a parent are not related in any way, caused mainly by forceful inheritance \cite{palomba2019impact}}&\parbox{4cm}{Replace inheritance with delegation}&\parbox{6.5cm}{Restructuring of code due to refactoring removes forceful inheritance, thereby preventing excess resource consumption}\\
    \hline
     temporary field &\parbox{3.5cm}{When an instance variable is set only for certain cases  \cite{Fowler}}&\parbox{4cm}{Remove unnecessary throws and unused parameters}&\parbox{6.5cm}{Refactoring results in removing temporary variables which act as additional fields ad consume CPU and memory in addition to the other variables}\\
    \hline
     speculative generality &\parbox{3.5cm}{Codes which are placed by programmers for anticipated future events \cite{samarthyam2016refactoring}}&\parbox{4cm}{Eliminate the smell by boxing objects to use static strings.}&\parbox{6.5cm}{As described above boxing the scalars to use the toString method is a waste of memory and CPU cycles.}\\
    \hline
    dead code &\parbox{3.5cm}{Obsolete code which was not removed \cite{gamma1995design}}&\parbox{4cm}{Parse through the methods to remove redundant code. }&\parbox{6.5cm}{Code which is no longer needed keeps calling the methods and allocate spaces in memory and consume CPU cycles. Refactoring removes this redundant code, thus stopping resource wastage}\\
    \hline
        duplicate code  &\parbox{3.5cm}{A code block which was copied in multiple classes rather than called through an object \cite{Fowler}}&\parbox{4cm}{Apply proper inheritance}&\parbox{6.5cm}{Removal of duplicate lines of code in multiple places will prevent CPU and memory from wastage}\\
    \hline
        long parameter  &\parbox{3.5cm}{When a method takes more than 5 parameters it is generally called to be having a long parameter \cite{samarthyam2016refactoring}}&\parbox{4cm}{Simplify the lambda using method reference}&\parbox{6.5cm}{Refactoring excessive number of parameters in a method will prevent caching at the beginning and this would remove the extra load on the CPU, however, memory usage will be increased}\\
    \hline
    long statement &\parbox{3.5cm}{A statement in a code, e.g. a switch statement containing too many cases \cite{refactoring.guru}}&\parbox{4cm}{Divide the long statement into smaller statements and establish proper communication between those. }&\parbox{6.5cm}{Refactoring prevents loading a long statement into memory which would otherwise consume excess memory resources.}\\
    \hline
     primitive obsession &\parbox{3.5cm}{The undesirable practice of using primitive types when representing an object. \cite{refactoring.guru}}&\parbox{4cm}{Use $StringBuilder$ which ensures that no locking and syncing is done, resulting in faster operation.}&\parbox{6.5cm}{Removes the use of obsolete string manipulation techniques like \textit{StringBuffer} which allows locking but no synchronization, thereby saving CPU and memory resource}\\
    \hline
    orphan variable &\parbox{3.5cm}{Variables that should be owned by another member class \cite{refactoring.guru}} &\parbox{4cm}{Extract all the variables to a class that should own them.}&\parbox{6.5cm}{Refactoring ensures that a variable is transferred to a class to which it should belong, thus preventing wastage of CPU cycles and memory spaces while doing this communication}\\
    \hline
     middleman &\parbox{3.5cm}{When a class is delegating almost all of its functionality to other classes \cite{refactoring.guru}} &\parbox{4cm}{Transfer functionality placed to the classes that they were mediating}&\parbox{6.5cm}{ Presence of such a delegation centered class will create extra overhead in terms of resource consumption which is prevented by refactoring}\\
    \hline
\end{tabular}
\caption{Characteristics and refactoring techniques of the analyzed software smells.}
\label{tab00}
\end{table*}
\hyphenpenalty 0

For automated smell detection and refactoring, we used \textit{jdeodrant}~\cite{fokaefs2011jdeodorant} and \textit{jsparrow}~\cite{jsparrow} for Java and \textit{pycharm} for Python \cite{gulabovska2019survey} applications to detect and refactor the code smells.

For each application, first, we compile and run the application without refactoring. In the process, we gather metadata in terms of CPU and memory usage. Second, we refactor them in two phases: in phase 1, we refactor all occurrences of one particular type of smell. In phase 2, we refactor multiple types of smells together to analyze the batch effect. Each application is run 14 times: seven times before refactoring and seven times after refactoring for each type of smell. We then take the average and standard deviation for the reported CPU and memory usage numbers. In total, we executed 6300 experimental runs for this study. 
Once all data is collected, we find the difference in CPU and memory usage before and after refactoring. Next, we normalize the differences in resource usage by the instance of each type of smell that was detected. This gives us the per smell impact of a specific type for each application. 

For method-level data collection, we use a tool called \textit{hprof} \cite{o2004hprof}. We record the execution path of the code and note the CPU and memory usage where the code smells are refactored. This allows us to collect information on resource usage precisely of the method which is refactored. As a result, we can relate the change in resource usage to refactoring. This is achieved by tracking resource usage via \textit{method\_id} which is unique to a method and assigned by the $hprof$ tool. Using hprof we collect resource usage data every 10 ms. For Python, we load the source codes in the \textit{pycharm} and compile the code. Afterwards, we apply specific workloads to test the resource usage before refactoring. When the applications are running, we execute the workload and collect the resource usage using \textit{logpid}. Next, we refactor the code smells in the same procedure discussed earlier and re-collect the resource usage data using the same workload. The workloads and experiments are detailed in the next section.

We conducted the experiments in cloud virtual machines which were created using \textit{kernel virtual machine (KVM)} over a bare metal server. The bare metal had 32 GB RAM, 8 core processors, and 2 TB persistent storage. We allocated a single core for every VM to make sure that the parallelization of processes does not affect the measurements. We ran each application 14 times. Every time a code smell was refactored, we executed the software in a new clean instance to eliminate the impact of previous run. 

For Java, we have selected 24 open source applications\footnote{The list of selected 24 Java applications and their details can be viewed on the GitHub page:  \href{https://github.com/asif33/batchrefactoring/blob/applications/java-apps.png}{https://github.com/asif33/batchrefactoring/blob/applications/java-apps.png}}
from Qualitas Corpus \cite{QualitasCorpus:APSEC:2010}, which is a dataset of 72 open-source Java applications. %
From the Corpus, we identified applications in five categories: code analyzers, code parsers, editors, email clients, and testing software. We selected applications which have more than 5000 lines of code (LoC), with at least 50 contributors in order to eliminate the risk of considering immature or a few developer contributed applications. 
For Python, we selected 7 open source applications\footnote{The list of selected 7 Python applications and their details can be viewed on the GitHub page: \href{https://github.com/asif33/batchrefactoring/blob/applications/python-applications.png}{https://github.com/asif33/batchrefactoring/blob/applications/python-applications.png}}
which have at least 5000 LoC and over 100 contributors.  
%
Next we explain the applications and workloads that were used in our experiments.


\subsection{Java: Applications and Workloads}
For Java applications, in order to understand the impact of refactoring on resource utilization, following workloads were run (clustered by application categories):


\textbf{Email clients.} The applications analyzed under this category are \textit{emf~ \cite{santos2020self}} and \textit{columba \cite{columba}}. Predefined email of size 70 bytes were sent using SMTP server \cite{riungu2011testing}. The emails were sent to 2920 users who were identified as mail readers. The average time to deliver an email is 3083.03 milliseconds with a median of 2847.3 milliseconds.

    
\textbf{Testing software.} Eclipse bug dataset \cite{zimmermANN2007predicting} is used as a workload, which contains data about six applications we analyzed in this category, namely \textit{jmeter \cite{halili2008apache}, findbugs \cite{ayewah2007using}, cobertura \cite{aarniala2005instrumenting}, emma \cite{liu2014java}, jstock \cite{QualitasCorpus:APSEC:2010},} and \textit{pmd \cite{rutar2004comparison}}.
We merged all classes and files into one large dataset which resulted in 24,642 LOC \cite{anvik2005coping}. The workload constituted of demo web services in Java which consisted of Java Server Pages (JSP), servlets, Enterprise Java Bean, and a database. The applications in the corpus were responsible for testing every conditionals and loop statements. 
    
\textbf{Editors.} We studied seven applications in this category, which are \textit{jedit \cite{wenzel2012isabelle}, jhotdraw \cite{savolskyte2004review}, antlr \cite{parr1995antlr}, aoi, galleon, batik \cite{QualitasCorpus:APSEC:2010},} and \textit{jruby \cite{nutter2014jruby}}. Multiple bots conducted activities in the editor such as typing, loading saved pictures, drawing simple shapes, and using various editor properties. The workload of each bot was 9.9 MB and a total of 109 virtual bots were used \cite{jacob2010actively}, \cite{myers1978controlled}. 
The total time for the workload of all bots was 180 seconds. 

\textbf{Project management.} The applications in this category include \textit{ganttproject \cite{cromar2013ganttproject}, xerces \cite{leung2004professional}, javacc \cite{dos2012compiler}, nekohtml \cite{QualitasCorpus:APSEC:2010}, log4j \cite{liu2014application},} and \textit{sablecc \cite{dos2012compiler}}. Multiple bots conducted project management activities \cite{armbrust2009berkeley}. Three sample projects was chosen: automated tender and procurement management, college management system, and resource monitoring system for ready-made garments. For each of the projects, bots checked whether the project management tool is available all the time.


\textbf{Parsers.} The applications considered in this category are  \textit{ant \cite{QualitasCorpus:APSEC:2010}, jparse \cite{james11jparse},} and \textit{xalan \cite{leung2004professional}}. The workload contained a set of incorrect and correct inputs \cite{duong2017multilingual}, \cite{ciric2005parsing}. The incorrect input is fed into the parser and ensured that the correct error code was returned by the parser. For the correct input, the expected Abstract Syntax Trees (AST) were described in a format that can be correctly parsed. The AST of the correct input was verified by a third party XML based parser considered to be bug free.


\subsection{Python: Applications and Workloads}

For Python applications, in order to evaluate the impact of refactoring on resource utilization, following applications and workloads were run:


\textbf{OpenStack.} OpenStack is a cloud platform providing the users with virtual machine instances which can be used for Software as a Service (SaaS), Platform as a Service (PaaS), or Infrastructure as a Service (IaaS) \cite{sefraoui2012openstack}. It is the most popular open source cloud platform in both academia and industry \cite{armbrust2009above}. To test the OpenStack source code, we compiled from source code and launched VM instances. The OpenStack processes including nova-compute were allocated to a single node and the resource consumption of that core were monitored.


\textbf{Sentry.} Sentry is a tool which reports and documents exceptions thrown by Python code russing at the back-end servers. Sentry runs in the background and acts as a central hub to monitor and report errors. Test case workloads provided by pytest were used for the experiments. We used the factory helper method to build data for the workload. The factory methods in \emph{sentry.testutils.factories} were available on all our test suite classes and stated in Sentry. We used the \emph{-k} option with \emph{pytest} to put workload on a single directory, single file, or single test depending on the scope of the code refactoring.


\textbf{Tensorflow.} To test \textit{Tensorflow} we use the $tf.distribute.Strategy$ to run the "2017 US Internal Migration" dataset and train the system. The database contained 80 years of data and it is used to train the \textit{Tensorflow} model to predict the internal migration trend for next 2 years. The size of the dataset was 3.2 GB large which contained detailed information regarding migration population, age, gender, occupation and economic conditions. 


\textbf{Tornado.} Tornado is a scalable framework with asynchronous networking library, primarily used to long-lived network applications. For Tornado, the inbuilt test suite was used to generate the data which will be used as a workload. The framework is synchronous, so the test results are completed when the method which is being tested returns. 

\textbf{Rebound.} Rebound is a popular tool used by software engineers. It is a command line tool written in Python that fetches all the solutions from stack overflow related to a problem. As a workload we called the \textit{sim.integrate(100.)} function which presents 100 pre-specified erroneous code blocks in rebound and relies on it to fetch the solutions from stack overflow. 

\textbf{Kivy.} Kivy is a Python library that is built over \textit{OpenGL ES 2} that allows rapid development of multi touch applications. The workload for Kiny was generated using its own module called \textit{recorder}, which allowed to replay keyboard events in a sample applications with Kiny running at the backend. A demo login page was launched with kivy which simulated clicking on a login button. the \textit{is\_click} option in \textit{recorderkivy.py} was set to true and the screen coordinates for the click was specified. Next the recorder was set to execute one click per second and this was repeated for 2 hours. This workload ensured that the critical code segments of kivy is called, a number of which also contained smells. 

\textbf{Falcon.} It is a WSGI library for building web APIs. The workload here mainly includes simulating requests to a WSGI client through \textit{class falcon.testing.TestClient(app, headers=None)[source]} class which is a contextual wrapper for \textit{simulate\_*()} function. This class will simulate the entire app lifecycle in a single call, which starts from lifespan and disconnecting process. This workload was repeated by passing the number of repetitions to the \textit{simulate\_request()} function. It was repeated 300 times and the CPU and memory usage were recorded. The same process was repeated after refactoring the falcon source code.

\begin{figure}[t]
    \centering
    \setlength{\belowcaptionskip}{-12pt}
        \includegraphics[width=0.5\textwidth]{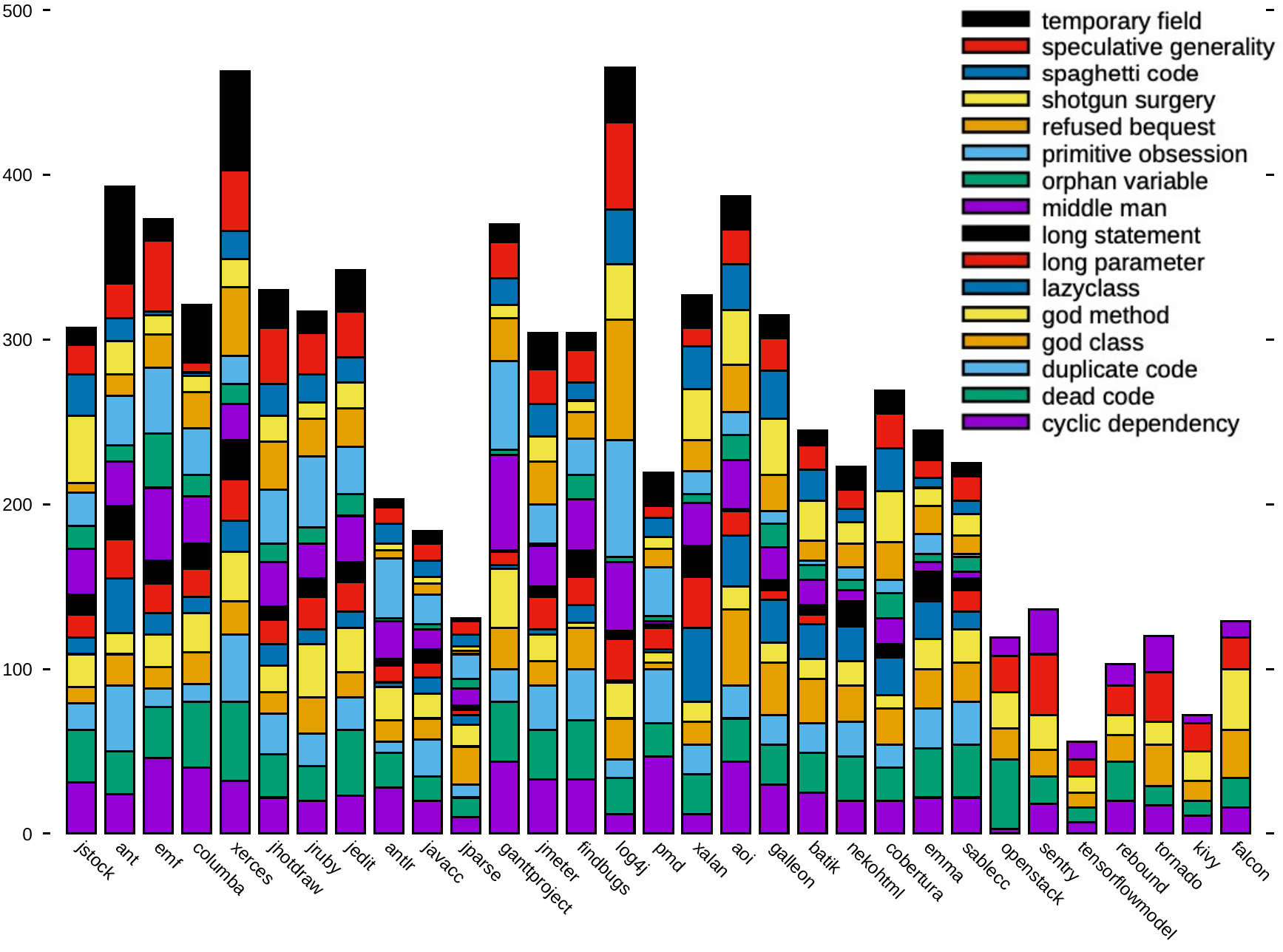}
    \caption{Code smell distribution across the 31 applications analyzed in this study.}
    \label{smellcount4343}
\end{figure}

\section{Results Analysis}\label{results}
\begin{figure*}[!htb]
    
    \begin{tabular}[t]{cc}

        \begin{tabular}{c}
        \smallskip
            \begin{subfigure}[t]{0.19\textwidth}
                
                \includegraphics[scale=1.3]{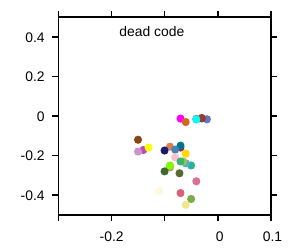}
               
            \end{subfigure}
            \begin{subfigure}[t]{0.19\textwidth}
              
                \includegraphics[scale=1.3]{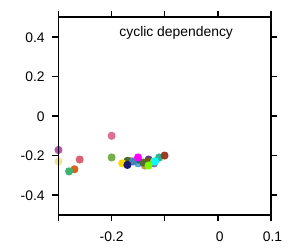}
              
            \end{subfigure}
            \begin{subfigure}[t]{0.19\textwidth}
              
                \includegraphics[scale=1.3]{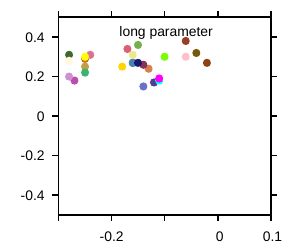}
            
            \end{subfigure}
            \begin{subfigure}[t]{0.19\textwidth}
             
                \includegraphics[scale=1.3]{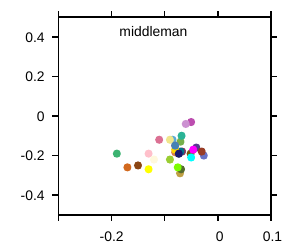}
               
            \end{subfigure}\\
            
            \begin{subfigure}[t]{0.19\textwidth}
                \centering
                \includegraphics[scale=1.3]{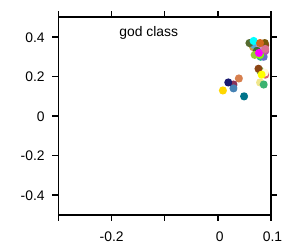}
               
            \end{subfigure}
            \begin{subfigure}[t]{0.19\textwidth}
                \centering
                \includegraphics[scale=1.3]{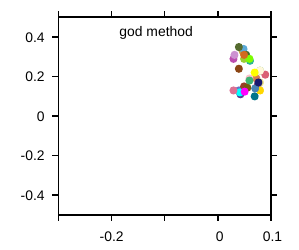}
             
            \end{subfigure}
            \begin{subfigure}[t]{0.19\textwidth}
                \centering
                \includegraphics[scale=1.3]{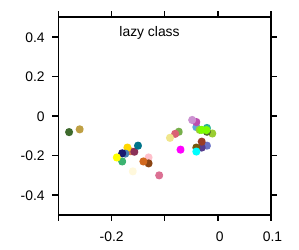}
             
            \end{subfigure}
            \begin{subfigure}[t]{0.19\textwidth}
                \centering
                \includegraphics[scale=1.3]{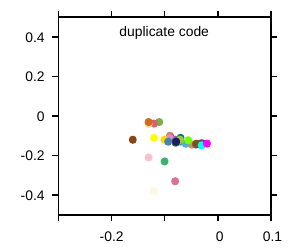}
             
            \end{subfigure}\\
            
            \begin{subfigure}[t]{0.19\textwidth}
                \centering
                \includegraphics[scale=1.3]{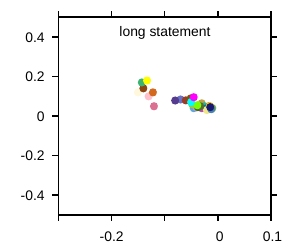}
              
            \end{subfigure}
            \begin{subfigure}[t]{0.19\textwidth}
                \centering
                \includegraphics[scale=1.3]{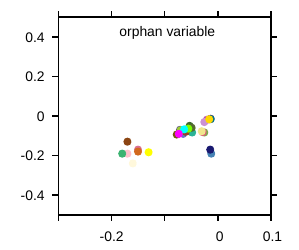}
             
            \end{subfigure}
            \begin{subfigure}[t]{0.19\textwidth}
                \centering
                \includegraphics[scale=1.3]{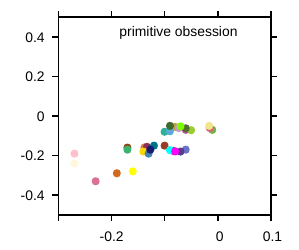}
               
            \end{subfigure}
            \begin{subfigure}[t]{0.19\textwidth}
                \centering
                \includegraphics[scale=1.3]{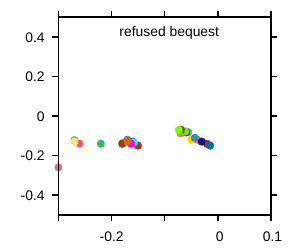}
              
            \end{subfigure}\\
            
            \begin{subfigure}[t]{0.19\textwidth}
                \centering
                \includegraphics[scale=1.3]{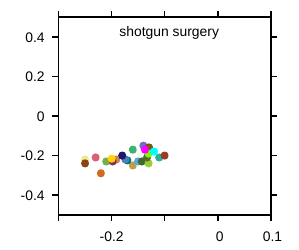}
              
            \end{subfigure}
            \begin{subfigure}[t]{0.19\textwidth}
                \centering
                \includegraphics[scale=1.3]{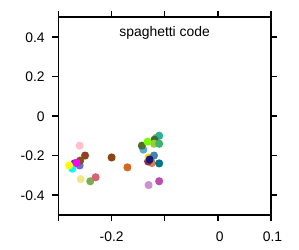}
               
            \end{subfigure}
            \begin{subfigure}[t]{0.19\textwidth}
                \centering
                \includegraphics[scale=1.3]{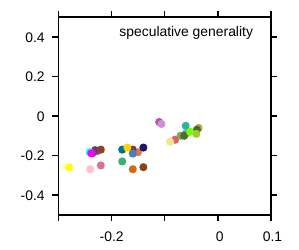}
              
            \end{subfigure}
            \begin{subfigure}[t]{0.19\textwidth}
                \centering
                \includegraphics[scale=1.3]{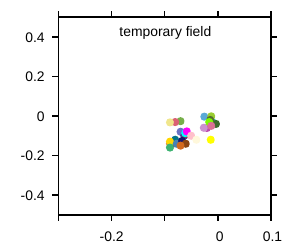}
               
            \end{subfigure}
        \end{tabular}
        
        &
        \begin{subfigure}{0.15\textwidth}
    \centering
    \includegraphics[width=1.1\linewidth]{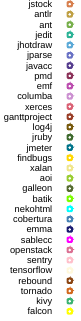}
\end{subfigure}
    \end{tabular}
    \caption{Normalized plots of impact of refactoring each code smell individually (per instance) on resource usage. X-axis: change in CPU usage (\%); Y-axis: change in memory usage (\%) of the application.}
     \label{fig:11}
\end{figure*}

In this section, we discuss the results of our experiments for both Java and Python applications. Figure~\ref{smellcount4343} shows the frequency of 16 analyzed code smell types across all studied applications. The distribution shows that while the number of smells differs between application no single smell is dominating. For example, \textit{Cyclic Dependency} code smell was prevalent in the highest numbers in Java source codes as we detected 725 instances of this smell as seen in the Figure. On the other hand, 259 instances of \textit{Orphan Variable} were detected. Given the assessment of smells distribution, we performed individual and batch refactoring of all applications, and we recorded the CPU and memory usage before and after the refactoring. Figure \ref{fig:11} shows the relative change in CPU and memory usage we observed. Here, we define relative change as the difference in CPU and memory usage between before and after refactoring.
The dataset for generating Figure \ref{fig:11} is provided  \footnote{\href{https://github.com/asif33/batchrefactoring/tree/scatter}{https://github.com/asif33/batchrefactoring/tree/scatter}} for reproducibility.
We note that in the case of Python applications our tools were able to detect and refactor on the following smells: dead code, cyclic dependency, long parameter, middleman, god method, and god class. Below, we summarize our findings for each type of smell.

\textbf{dead code}: We know that dead code is a code that is either redundant, because the results are never used, or is never executed. Since the results are never used but the code is getting executed, it is common to expect that it leads to CPU and memory waste. In cases when the code is not executed, it can still have adverse effect due to adding code bloat. Our results confirm that the CPU usage can be improved by removing dead code smells. Dead code makes the runtime footprint larger than it needs to be, thereby consuming excess resource in terms of CPU and memory which can be critical for large scale data center applications like \textit{OpenStack} as studied in this research. 

\textbf{cyclic dependency}:  This smell can cause a domino effect on the code when a small change in one module quickly spreads to other mutually recursive modules. The smell caused infinite recursion in 134 instances where it was found. In 63 instances, it resulted in memory leaks in java by preventing garbage collectors to deallocate memory. The extent of the impact of this smell is also dependent on the type of software as a similar type of application is found to behave similarly. Analysis of the refactored code shows that the refactoring process eliminates the enhanced loops in most parts of the software, thereby improving resource usage. The enhanced loop traverses each loop one by one, thereby requires increased CPU even when traversal of the entire array may not be required. The refactoring tools address such cases and removes the enhanced loops. 

The removal of unwanted loops results in loop unrolling which is observed in the refactored code of both Java and Python datasets. The loop unrolling reduces CPU and memory consumption by removing loop overhead. At the same time, loop control instructions and loop test instructions are eliminated, so the resource required to conduct those activities are freed. The total number of iterations are reduced to improve resource efficiency. As seen in the figure, in all cases of Python and Java, removal of cyclic dependency code smell is seen to improve resource utilization performance.

Considering $jstock$, it is seen that the refactoring of cyclic depend code smells results in 5.89\% CPU and 6.16\% memory, with a standard deviation of 0.26 and 0.49 respectively. In other cases of Java dataset, improvements are noticed as well. When we consider the dataset of Python, we see that removal of cyclic dependency code smell decreases CPU and memory consumption for a specific workload of \textit{tensorflowmodel} by 0.33\% and 0.21\% respectively for each smell refactored. 

\textbf{long parameter}: For long parameter code smell, it is seen that out of the 24 applications, all are showing positive memory change and negative CPU change, meaning memory usage degraded after refactoring the software smell.  When we look towards refactoring, for example in $jhotdraw$ the tool used “Introduce Parameter Object” refactoring. 

If we consider an example of a long parameter smell found in $openstack$ which is a software of our experimental dataset of Python, we noticed that a method with many parameters is refactored where the parameters are distributed to three methods preserving the functionality. Although the above segregation is a better way to provide useful and reusable classes, it is causing unboxing of the parameters from one method to 3 methods. If one method contained all parameters then, all those parameters could have been cached at the beginning and it does not require loading into memory multiple times. However, this would provide an extra load on the CPU as the parameters which were not required at the initial stage of polygon formation would still be called. Refactoring it in the mechanism described above will break the concatenation, hence preventing caching. On the other hand, the prevention of caching all the parameters at the very beginning will cause excess memory to be used. Hence, refactoring this smell for the above 2 types of applications will reduce CPU usage but worsen memory usage. 

Although, the modularization of code improves readability, it will worsen memory usage as more instructions need to be loaded into memory. Similar behavior applies to the remaining 3 applications which are showing these traits. For $openstack$ we notice that the CPU utilization reduces by 7.9\% which I significant compared to others. It must be stated that the number of smells of the long parameter in $openstack$ was found to be 40, significantly higher than the same smell being found in other apps. This large number of smells may have contributed to the improvement of CPU usage.

\textbf{middleman}: Elimination of middle man smells contributed to the improvement of CPU and memory usage. The most improvement is seen in $ganttproject$ with CPU and memory usage reductions of \textit{0.61\%} and \textit{0.29\%} respectively. There were \textit{58} instances of middleman code smell in the \textit{ganttproject}, this resulting in a significant performance improvement. Also, the \textit{ganttproject} is a CPU-intensive project occupying a significant percentage of CPU when running, thus yielding greater change in CPU than memory. For Python dataset, $sentry$ had the maximum number of middleman code smell which was detected. When the 27 code smells in $sentry$ were refactored, per smell improvement in CPU and memory was 0.44\% and 0.13\% respectively for every smell refactored.

\textbf{god class}: In the list of applications that were refactored it is seen that the extract class refactoring mechanism caused the resource usage to worsen \cite{alkharabsheh2021exploratory}. Refactoring this smell involves a large class is separated into multiple smaller classes, each with lesser responsibilities, hence extra time and resources are required for inter-class communication, as a result, CPU and memory usage increases. Further analysis shows that the inter-class communication increased as large classes were extracted into multiple small classes. We tool all the new methods which are created and found the average lines of code in those. In most cases, we see that usually 16.14 lines of code trigger and complete operations on a variable or object, based on slicing a new method needs to be made with those. From a software engineering perspective, such large volumes of extraction are desirable, however from the standpoint of the resource usage, such granular segregation may cause huge volume of context switching and inter-method communications, which may add high volume of overhead.

\textbf{god method}: The behavior of graph for god method is similar to that of god class. All values are positive which shows that refactoring the god method code smells increase resource usage. Besides, the normalized increase is quite high for the god method compared to other kinds of code smells. To refactor the god method, the extract method mechanism is used. So a large method is broken down into multiple smaller methods, which increases inter-method communication.

\textbf{lazy class}: The same behavior is seen for refactoring lazy class smell where each category of applications is showing similar behavior. One exception is that JRuby is located very close to the group of document editors. The reason is that the number of lazy class smells of JRuby is only 9. Similar resource consumption changes are seen for the group of editors where the number of smells ranges from 9-13 for all the applications. Hence for $lazy$ $class$ the number of smells is proportional to the impact on resource usage.

\textbf{duplicate code}: Similar behavior is seen for refactoring duplicate code smell. It is seen that the apps belonging to the same category are behaving similarly, emphasizing the fact that similar types of apps have the same impact when the code smell is refactored. $Ant$, $xalan$, $maven$, and $xerces$ are found to show significant improvement in CPU resources after refactoring. Analysis of the code in $xerces$ shows that it parsed the XML documents and placed the variables in those in a list, reiterating through it multiple times. 

\textbf{long statement}: Results of the long statement are seen to be in line with the results of the long parameter. The CPU change after refactoring is lowered whereas the memory usage increases. However, although the increase varies differently for a different group of applications, the parser category shows high usage of memory compared to the other categories.

\textbf{orphan variable}: Similar category of applications are seen to behave similarly in terms of change in resource usage when the orphan variable is refactored. As a result, it can be stated that the category of applications can be used to group the impact of refactoring the code smell. The email clients namely $emf$ and $columba$ are seen to have the maximum impact of refactoring this code smell.

\textbf{primitive obsession}: After refactoring primitive obsession code smell and normalizing with the count of smells, it is seen that for primitive obsession the change in resources data can be used to group the applications by category. One of the rules which are used by the refactoring tool is to replace $StringBuffer$ with $StringBuilder$. It is recommended to use $StringBuilder$ because no locking and syncing is done. Hence, it's faster. When running programs in a single thread, which is generally the case, $StringBuilder$ offers performance benefits over $StringBuffer$. 

\textbf{refused bequest}: We see that the impact of automated refactoring is higher for the group of code analyzer apps than the others. This is because the testing apps loaded the source code in memory to run the tests. The presence of unused methods and variables in the code which is loaded into memory resulted in excessive resource usage by the applications. On average, after refactoring 0.284\% of CPU and 0.147\% of memory were reduced for each refused bequest smell refactored.

\textbf{shotgun surgery}: For log4j, it is seen that refactoring the shotgun surgery significantly contributed to improving memory resources by 7\%. Refactoring this code smell also improved the unpredictability and efficiency of the generated random values. Simplification of the data structures occurred in 21 of the cases of refactoring, a high percentage of 61.76\% where this refactoring was done, thus simplifying the code significantly. As most of the loops were used to read and load the logs in memory, simplifying it meant that less memory will be required for loading. 

\textbf{spaghetti code}: The $jruby$ application had the highest impact of refactoring the spaghetti code smell. The number of spaghetti codes detected in this application is 57, which is higher than any application in the list. This resulted in more loc being refactored and greater change in resource usage before and after refactoring. One of the rules of refactoring spaghetti code replaced the $concat()$ method on Strings with the + operator. It should have slight performance benefits if the size of the $concat()$ is large. Another rules replaced $length()$ or $size()$ with $isEmpty()$. This rule should provide performance advantages since $isEmpty()$ time complexity is $O(1)$ whereas $length()$ and $size()$ can have a time complexity of $O(n)$. 

\textbf{speculative generality}: It is seen that the code parser category showed the highest change in CPU and memory utilization for speculative generality. This category of applications has 76 cases of speculative generality and non-normalized CPU usage improved by 4.63\% and memory improved by 1.47\% due to refactoring of the smells. This increase is mainly due to the removal of excess code that was added but not called in the system. These codes kept using heap memory and used CPU for basic non-required computations.


\textbf{temporary variable}: Given the lower number of smells detected for this smell type in the applications, the grouping of applications in the plots based on category implies that the smell is having an impact on the resource change. Also, the refactoring does not keep temporary fields, thus making those final, leading to improvement in resource consumption.

\subsection{Impact of Batch Refactoring}
\label{combined}

\begin{figure}
  \centering
  \begin{tabular}{@{}c@{}}
    \includegraphics[width=1.0\linewidth,height=100pt]{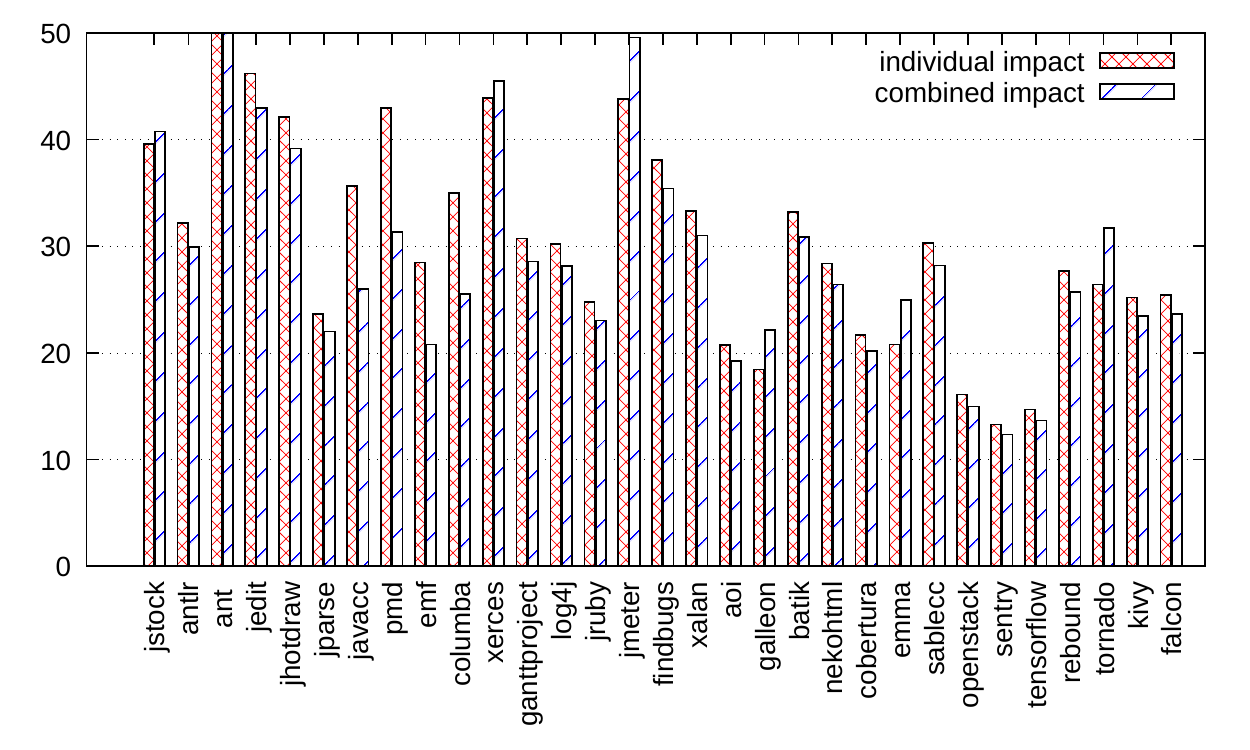} \\[\abovecaptionskip]
    \small (a) impact on CPU
  \end{tabular}
\vspace{-1\baselineskip}
  \begin{tabular}{@{}c@{}}
    \includegraphics[width=1.0\linewidth,height=100pt]{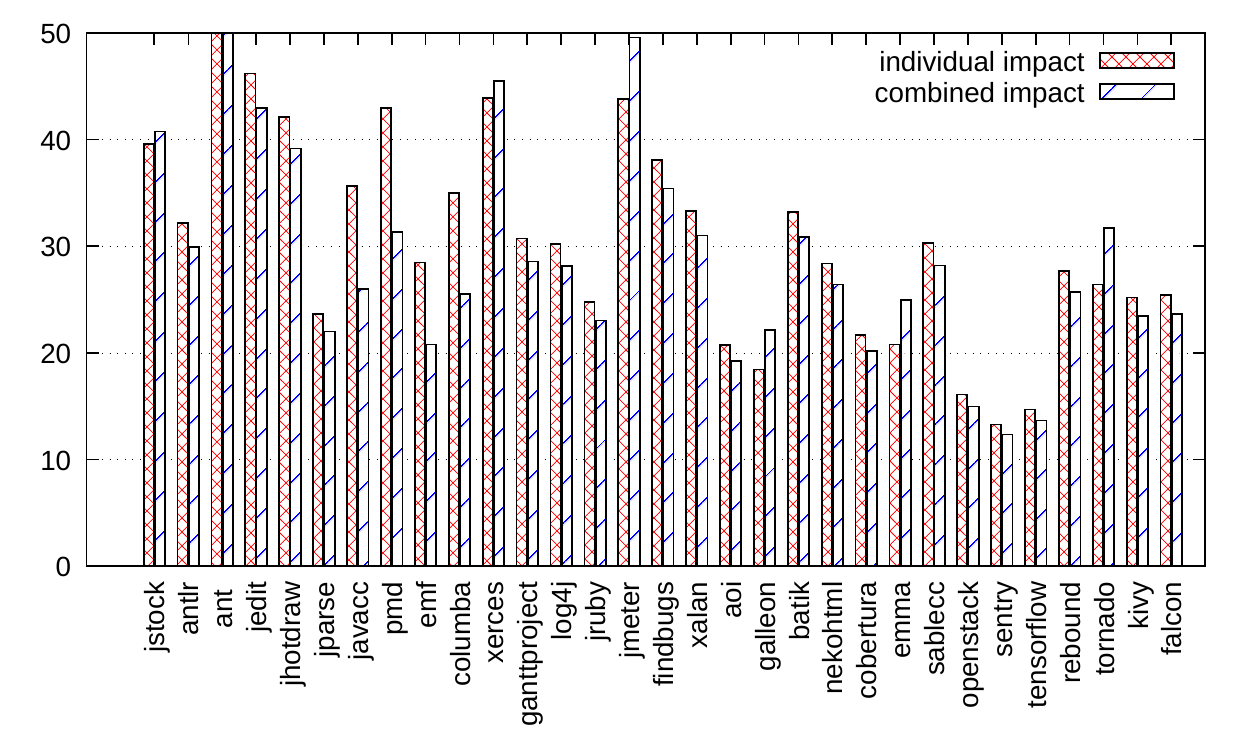} \\[\abovecaptionskip]
    \small (b) impact on memory
  \end{tabular}
\vspace{1\baselineskip}
  \caption{Combined impact of smell refactoring which worsen resource usage.}\label{worsen}
\end{figure}

In the previous section, we did a benchmark by analyzing the impact of individually refactoring the smells on resource usage. Although it helped us determine a benchmark, however, in real life no software occurs with a single type of smell only. Hence it is very important to see the combined impact of smell refactoring on resource usage. Also, we want to see whether the combined impact adds up to the individual impact of refactoring different smell types since this will ensure that the change in resource usage is caused by the code smell refactoring. With this requirement in mind, we proceed to refactor the smells in batch as shown in Figure \ref{all}, \ref{improve}, and \ref{worsen}. The dataset of the Figures can be viewed from the link \footnote{\href{https://github.com/asif33/batchrefactoring/tree/batch-refactor}{https://github.com/asif33/batchrefactoring/tree/batch-refactor}}.

\subsubsection{Refactoring all code smells. \\}

\begin{figure}
  \begin{tabular}{@{}c@{}}
    \includegraphics[width=1.0\linewidth,height=100pt]{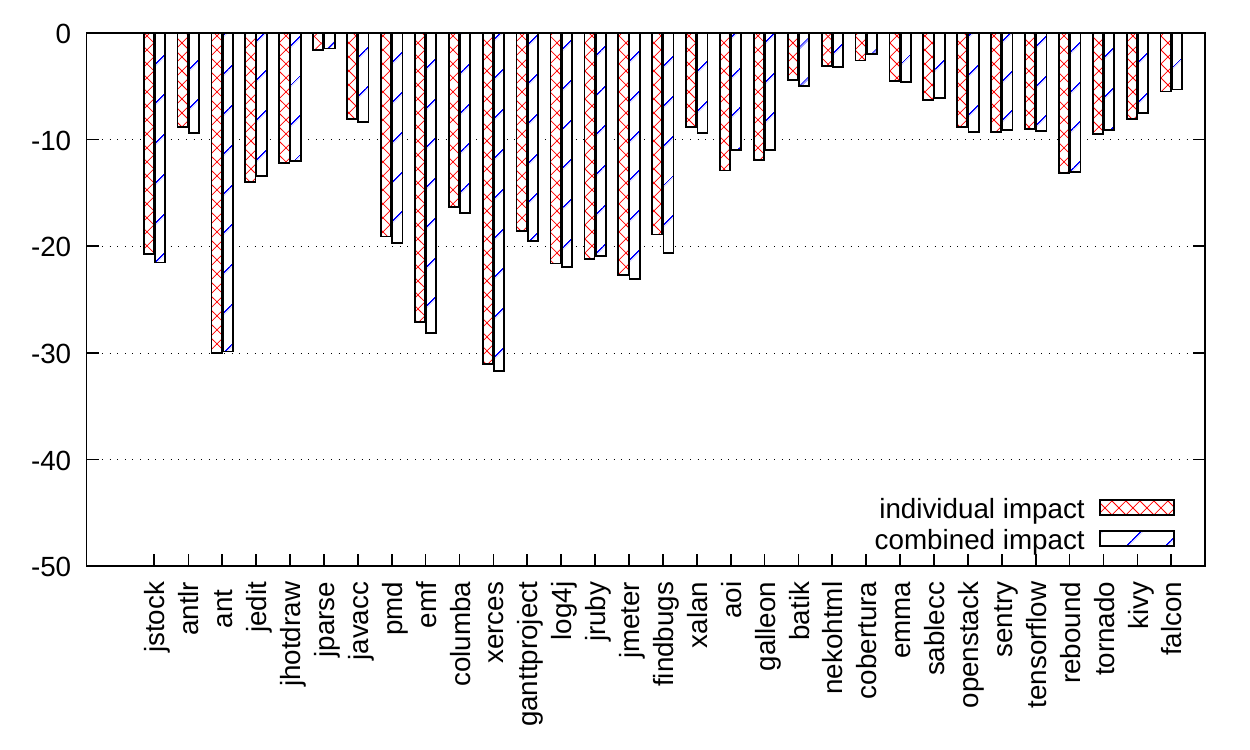} \\[\abovecaptionskip]
    \small (a) impact on CPU
  \end{tabular}
\vspace{-1\baselineskip}
  \begin{tabular}{@{}c@{}}
    \includegraphics[width=1.0\linewidth,height=100pt]{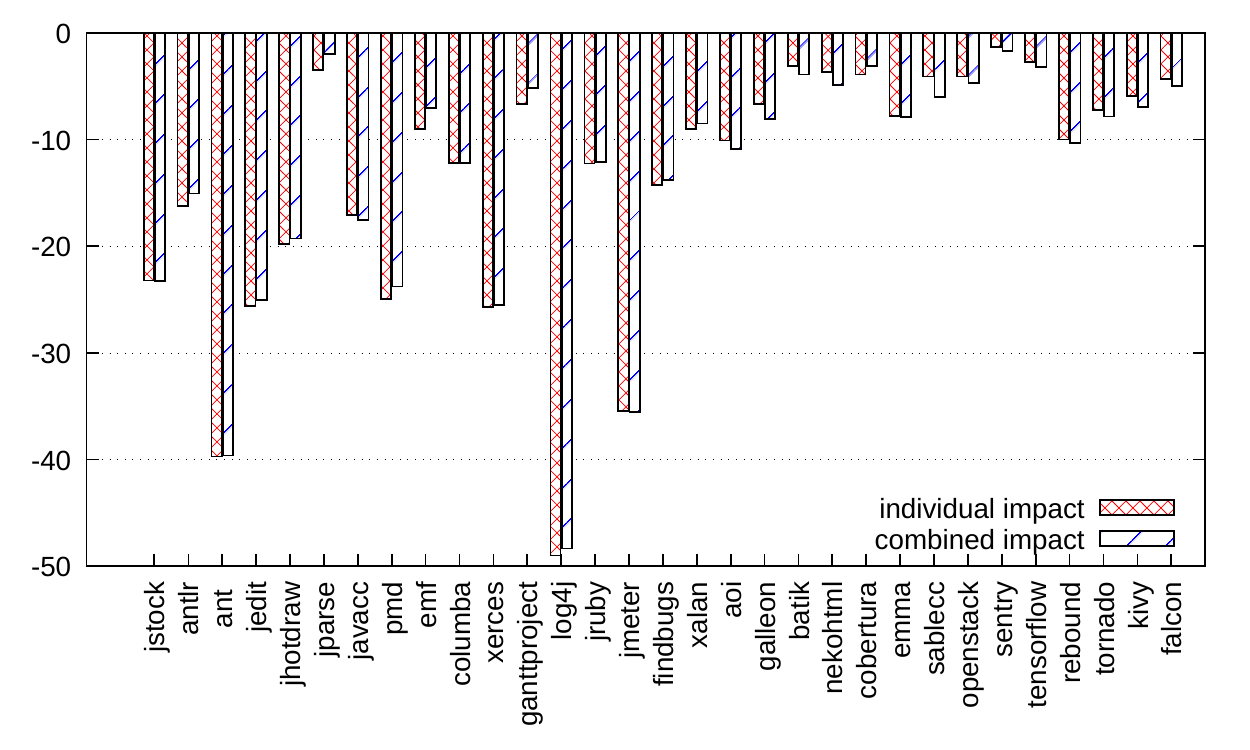} \\[\abovecaptionskip]
    \small (b) impact on memory
  \end{tabular}
\vspace{1\baselineskip}
  \caption{Combined refactoring impact of all smells considered in this research.}\label{all}
\end{figure}

Here we refactored and analyzed the impact of the 16 smells altogether. In this section, we provide the findings in terms of CPU and memory usage. We analyze the impact on CPU and memory separately. Figure \ref{all} shows the impact of refactoring those smells.

\textit{Impact on CPU:}
Combined refactoring of all the smells, irrespective of impact can provide useful information as to whether those smells improve resource usage or worsen those. It is seen that although the individual impacts of performance degrading smells is significant, refactoring all the 16 smells in 24 applications resulted in improvement of the resource usage overall since the type of resource usage improving smells were larger compared to those which worsen performance. From the CPU perspective, it is seen that the total CPU usage of ant improved by 30.01\% which is significant and desirable. At the same time, the least percentage improvement was seen in $Javacc$ which is 8.10\%. It is seen that the percentage improvement of CPU is greatly influenced by the presence of the numbers of various types of smells.

\textit{Impact on memory:}
A similar pattern is seen for memory consumption where the usage improves after refactoring the 16 smells studied in this research where $ant$ showed the highest improvement of 39.70\%. The lowest change in CPU usage was seen for $jparse$ with a 3.50\% improvement. Again the increase can be credited to the total instances of the various types of smells found in the un-refactored code.

\subsubsection{Refactoring code smells that increase resource usage.\\}

\textit{Impact on CPU:}
It is seen that refactoring god class, god method, and feature envy negatively impact performance when refactored. Upon analysis of the normalized graph for god class and god method, it is seen that per smell impact of god class is found to be around 0.22\%-0.50\%, whereas for god class it is 0.20\%-0.22\%, indicating that a software engineer who is focusing on refactoring and has optimizing resource usage in mind should avoid refactoring god classes and god methods.

At the same time, it is seen that $ganttproject$ suffers from the largest percentage increase of CPU usage which is undesirable. In total $ganttproject$ had 61 occurrences of smells of god class and god method, refactoring which greatly impacted to resource usage degradation of 16.30\%. On average the total degradation of CPU usage after refactoring the smells for 24 applications is found to be 7.79\%. Upon refactoring the individual smells and adding the total change in CPU usage, we get similar values to refactoring them altogether.

\textit{Impact on memory:}
Refactoring god class and god method worsened memory usage as well for the 24 applications. Log4j had the highest degradation of memory consumption which is 19.50\% when the concerned smells were refactored altogether. Also, refactoring those individually and adding up the values resulted in total memory consumption of 20.01\%, which is only 0.51\% greater than combined refactoring, ensuring that the results are adding up to individual impact and hence they are consistent. The large number of smells of god classes and god methods which sum up to 100 instances of smells being present in the code leads to a large volume of refactoring done in the code by the implementation of extract class and extract method refactoring procedures. This contributed to the large distortion of memory usage before and after refactoring. Overall, we see that individual refactoring impacts add up when the refactoring is done combined procedure. The mean deviation in CPU is 0.64\% and the mean deviation in memory is 1.47\%.

\subsubsection{Refactoring code smells that decrease resource usage.\\}

\begin{figure}
  \centering
  \begin{tabular}{@{}c@{}}
    \includegraphics[width=1.0\linewidth,height=100pt]{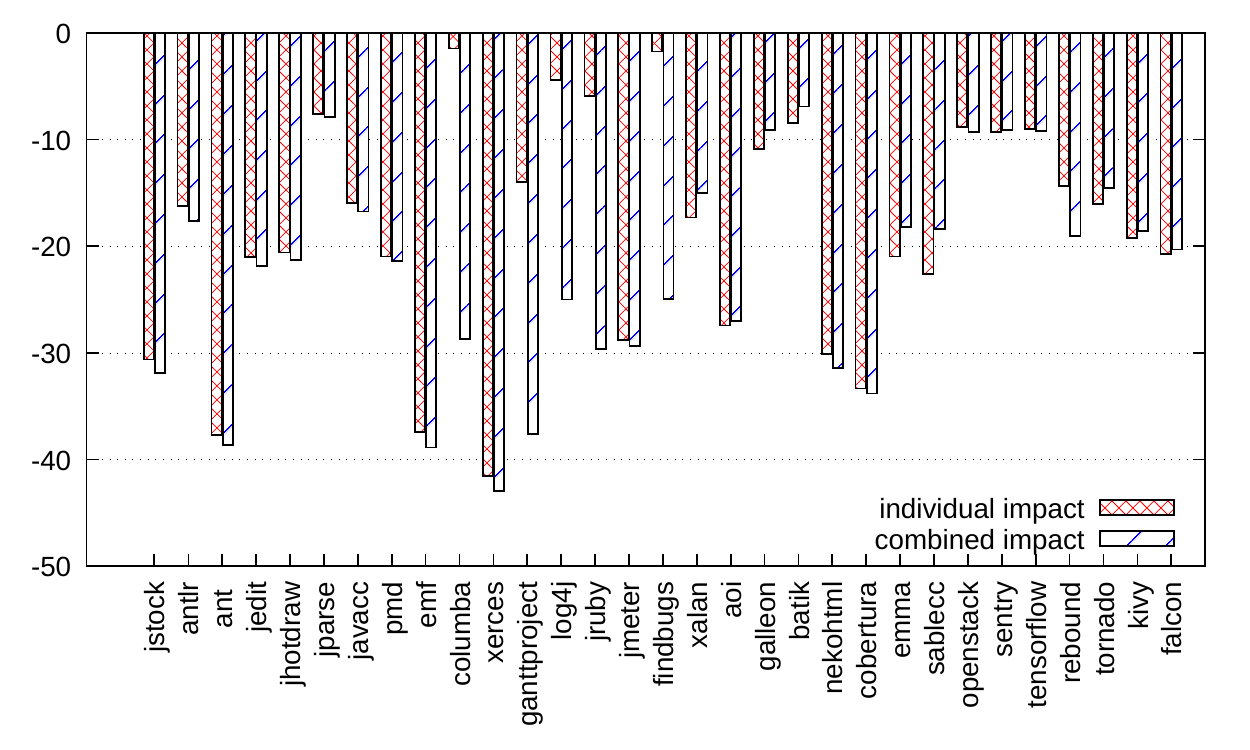} \\[\abovecaptionskip]
    \small (a) impact on CPU
  \end{tabular}
\vspace{-1\baselineskip}
  \begin{tabular}{@{}c@{}}
    \includegraphics[width=1.0\linewidth,height=100pt]{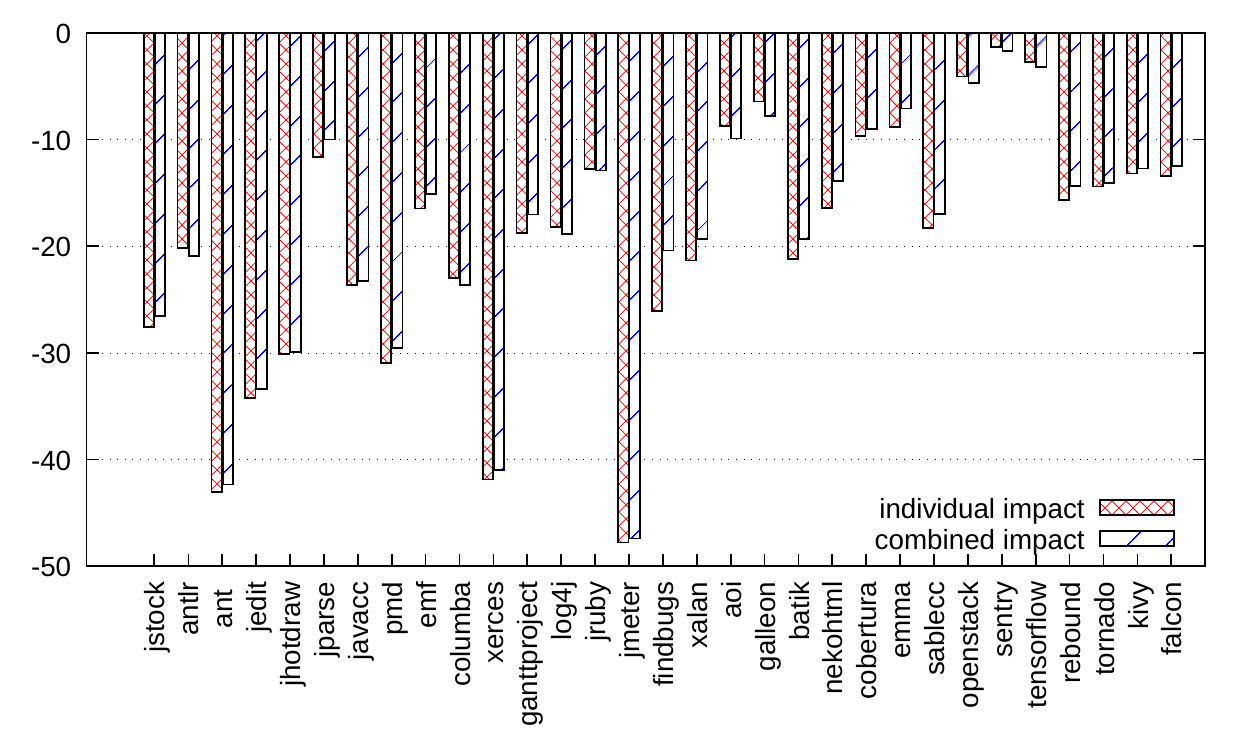} \\[\abovecaptionskip]
    \small (b) impact on memory
  \end{tabular}
\vspace{1\baselineskip}
  \caption{Combined impact of smell refactoring which improve resource usage.}\label{improve}
\end{figure}

This section states the impact of auto-refactoring all the smells at once which positively impacts resource usage. Similar to the last section, we highlight the impact on CPU and memory separately and analyze the consistency as shown in Figure \ref{improve}.

\textit{Impact on CPU:}
We analyzed refactoring which smells consistently improved performance for our dataset of 24 apps in this study and found that \textit{cyclic dependency, duplicate code, dead code, primitive obsession, speculative generality, shotgun surgery, long parameter, middle man, refused bequest, orphan variables, long statements}, and \textit{temporary fields} were seen to meet our condition. We proceeded to refactor the aforementioned smells altogether and determine the total change in CPU and memory when they are refactored combinedly. Out of the 24 applications, $columba$, $log4j$, and $jruby$ gave errors when the smells which improved performance individually were refactored together. As shown in Figure \ref{fig:11} for CPU, it is seen that the range of percentage improvement of CPU stretched from 7.6\% for $jparse$ till 37.70\% for $ant$. We proceeded to sum the impact of refactoring those smells individually for comparison purposes. It is seen that combining impact stretched from 7.86\% to 38.87\%. Upon calculation of the differences, it is seen that combining all the smells whose refactoring improves performance show consistent behavior to refactoring them individually and adding the values. The difference in the values ranged from 0.26\% to 1.46\% which were seen for $jparse$ and $emf$ respectively. The mean deviation is 0.61\%.

\textit{Impact on memory:}
The memory usage is impacted significantly with a range of 25.47\% and 47.77\% for the apps and an average improvement of 28.63\%. It is seen that $jmeter$ shows the maximum improvement whereas $emf$ shows the minimal effect on memory. Further analysis shows that when the smells are refactored in $jmeter$, the volume of spatial locality is increased due to the rearrangement done to it. Also, compression is conducted by dissolving longer parameters which result in smaller and smarter formats. Finally, temporal localities are increased by refactoring smells like a refused bequest, shotgun surgery, and speculative generality which cache trashing, hence reducing memory usage. Reduce and reuse refer to techniques that minimize memory operations with the temporal locality that reduce cache fetches. This is accomplished by reuse of data still in the cache by merging loops that use the same data with a mean deviation of 0.64\%.

Based on the findings of this section, we use the experimental data to deploy train and test machine learning techniques to predict the impact of the smells based on the detection and before they are refactored.

\begin{table}
\begin{tabular}{ l c } 
 \hline
 \textbf{Smell} & \textbf{Mean difference} \\ \hline 
 cyclic dependency & 0.070 \\
 \hline 
 dead code & 0.095 \\
 \hline 
 middleman & 0.045\\
 \hline 
 long parameter & 0.055\\
 \hline 
 god class & 0.060\\
 \hline
 god method & 0.095\\
 \hline
\end{tabular}
\label{general}
\caption{Generalized impact of code smell refactoring}
\end{table}
\subsection{Predicting Resource Utilization Impact}
\begin{table*}[h]
\begin{tabular}{l c c c c c c c c c c}
\hline
\multicolumn{1}{l}{\textbf{}} & \multicolumn{2}{c}{\textbf{Linear regression}} & \multicolumn{2}{c}{\textbf{Polynomial regression}}& \multicolumn{2}{c}{\textbf{Lasso regression}}& \multicolumn{2}{c}{\textbf{Random forest}}& \multicolumn{2}{c}{\textbf{ANN-regression}}                                            \\ \hline
\textbf{code smell}  & mse & rmse & mse & rmse & mse & rmse & mse & rmse & mse & rmse \\ \hline
cyclic dependency & 1.50 & 1.78 & 1.41 & 1.66 & 0.73 & 0.89 & 0.53 &0.71 & 0.43 & 0.62 \\ \hline
god class & 1.85 & 2.01 & 0.63 & 1.03 & 0.66 & 0.89 & 0.47 &0.66 & 0.31 & 0.37 \\ \hline 
god method & 0.84 & 0.96 & 0.76 & 0.81 & 0.62 & 0.70 & 0.47 & 0.56 & 0.25 & 0.43 \\ \hline
dead code & 1.42 & 1.59 & 0.32 & 0.51 & 0.32 & 0.50 & 0.29 & 0.46 & 0.22 & 0.32 \\ \hline 
long parameter & 1.52 & 1.61 & 0.41 & 0.51 & 0.33 & 0.49 & 0.21 & 0.36 & 0.19 & 0.22 \\ \hline 
middleman & 1.67 & 1.98 & 0.81 & 1.12 & 0.71 & 0.98 & 0.44 &0.86 & 0.21 & 0.28 \\ \hline 
\end{tabular}
\caption{Results of multivariate regression analysis for memory.}
\label{predictionresults}
\end{table*}
In this section, we proposed an approach based on machine learning (ML) using different metrics of the software and the number of code smells detected to predict the resource consumption changes due to code smell refactoring. We find that the selection of relevant software metrics as features plays an important role in the performance of the ML algorithms. Our dataset was created using the benchmarking procedure discussed earlier in this paper. We used a genetic algorithm for feature selection and find that for all the algorithms, the performance is best is improved for a certain combination of relevant features. 

Our experiments included four machine learning algorithms namely linear regression, polynomial regression, lasso regression, random forest regression, and ANN-regression. The ANN-regression model achieved the best performance as shown in Table \ref{predictionresults}. The table shows the regression results for the 6 code smells which were detected in both Java and Python applications. The rest of the results can be found  \footnote{\href{https://github.com/asif33/batchrefactoring/blob/applications/regression-data-remaining-smells.png}{https://github.com/asif33/batchrefactoring/blob/applications/regression-data-remaining-smells.png}}.We see that the regression algorithms have a high potential for predicting resource consumption impact by refactoring code smells.

Despite determining the CPU impact for individual and batch refactoring of code smells, software engineers want to identify this impact even before they conduct refactoring. This will enable them to decide which software code smells to address during the refactoring phase to save time and engineering effort. Here, we used the data of bench-marking to predict resource usage change via regression analysis. In this regard, we identified some features of the code smells such as lines of code with smells, weighted method per class, FanIn, FanOut, and category of applications, number of smells which impact resource usage. We applied a $Naive$ approach which involved taking the mean of normalized CPU and memory of all applications except for the target application. Next, we used the mean to predict the resource usage of the target app by considering the identified features. We proceeded to calculate the \textit{Mean Squared Error (MSE)} for both the CPU and memory. The \textit{MSE} for CPU was found to be $0.02216$ and for memory, it was $0.03165$.

After the $Naive$ approach, we applied linear regression to predict the CPU and memory usage. We used features of the code smells to predict the impact on resource usage. The first part of the exercise considered the impact prediction of individual features. Next, we conducted multivariate regression analysis which provided the impact of each independent variable on the dependent variables. We see that the $MSE$ is minimum when we combined all independent variables in the multivariate approach. This result is desirable and low $MSE$ indicates that we can predict the change resource usage for refactoring code smells in apps even before we do the refactoring. The $MSE$ for multivariate regression for CPU was $0.01161$ and for memory, it was $0.02011$.

The estimate is the average impact per refactored smell on CPU and memory by a specific feature when all other features remained constant. This is used as a coefficient in the regression formula. The ratio of the estimate and standard error provided us the $t-value$. Positive $t-values$ were observed for all 16 smells. \textit{Adjusted r-squared} was found to be $0.891$ and $0.833$ for CPU and memory respectively, indicating that the model can explain \textit{89.1\%} of the variations in the training data set. Although regression analysis gives us promising results, there is a need to conduct a detailed future study in prediction of code smells impact on resource usage.

\section{Related Work}
\label{relatedwork}
Automated batch refactoring techniques are known to significantly improve overall software quality and maintainability, but their impact on resource utilization is not well studied in the literature. Oliveira et al. conducted an empirical study to evaluate nine context-aware Android apps to analyze the impact of automated refactoring of code smells on resource consumption \cite{oliveira2018empirical} of Android applications. They studied three code smells, namely god class, god method, and feature envy. They found that for the three smells, resource utilization increases when they are refactored. Although their findings are useful, it is limited to the analysis of three code smells only. At the same time, the importance of analyzing the impact of batch refactoring code smells on software resource usage was not considered.


To understand the relationship between Android code smells and nonfunctional factors like energy consumption and performance, Palomba et al. \cite{palomba2019impact} conducted a study with nine Android-specific smells and 60 Android applications. Their results showed that some smell types cause much higher energy consumption compared to others and refactoring those smells improved energy consumption in all cases. Although the results are consistent with our findings, the authors only addressed the individual impact of nine code smells and the analyzed smells were specific to Android applications.

The impact of multiple refactoring on code maintainability, also known as batch refactoring, was explored by Bibiano et al.~\cite{8870183}. They argue that removing an individual code smell in a code block increases the tendency of introducing new smells by 60\%. Therefore, the importance of analyzing the combined and complex impact of refactoring code smells in a batch rather than individual smells is proposed.  Besides maintainability, it is also essential to study the effect of batch refactoring on the resource usage of the application.

Park et al. investigated whether existing refactoring techniques support energy-efficient software creation or not \cite{park2014investigation}. Since low-power software is critical in mobile environments, they focused their study on mobile applications. Results show that specific refactoring techniques like the \textit{Extract Class} and \textit{Extract Method} can worsen energy consumption because they did not consider power consumption in their refactoring process. The goal was to analyze the energy efficiency of the refactoring techniques themselves, and they stated the need for energy-efficient refactoring mechanisms for code smells.

Platform-specific code smells in High-Performance Computing (HPC) applications were determined by Wang et al. \cite{wang2014platform}. AST-based matching was used to determine smells present in HPC software. The authors claimed that the removal of such smells would increase the speedup of the software. The assumption was that specific code blocks perform well in terms of speedup in a given platform. However, the results show that certain smell detection and refactoring reduced the speedup, thus challenging the claims and showing the importance of further research in this area.

Pérez-Castillo et al. stated that excessive message traffic derived from refactoring god class increases a system’s power consumption \cite{perez2014analyzing}. It was observed that power consumption increased by 1.91\% (message traffic = 5.26\%) and 1.64\% (message traffic = 22.27\%), respectively, for the two applications they analyzed. The heavy message-passing traffic increased the processor usage, which proved to be in line with the increase in the power consumption during the execution of those two applications. The study was limited to only god class code smell. However, a detailed analysis is required to determine the impact of code smell refactoring on resource consumption.

An automatic refactoring tool that applied the \textit{Extract Class} module to divide a god class into smaller cohesive classes was proposed in \cite{fokaefs2011jdeodorant}. The tool aimed to improve code design by ensuring no classes are large enough, which is challenging to maintain and contains a lot of responsibilities. The tool refactored code by suggesting \textit{Extract Class} modifications to the users through a User Interface. The tool was incorporated into the Eclipse IDE via a plugin. The authors consulted an expert in the software quality assessment field to give his expert opinion to identify the effectiveness of the tool. Results show that in 12 cases (75\%), the evaluator confirmed that the classes suggested being extracted indeed described a separate concept. According to the expert, two of these classes could be extracted and used as utility or helper classes. However, the effect of such refactoring on resource usage of the software was considered to a limited extent.

The results showed that refactoring smells by automated tools like JDeodorant and JSparrow have widely varying impacts on the CPU and memory consumption of the tested applications based on the specific smell types. We presented each smell's resource utilization impact and discussed the potential reasons leading to those effects.

\section{Conclusion}
\label{conclusion}
In this paper, we evaluated the impact of batch refactoring 16 code smells on the resource usage of 31 open-source Java and Python applications. We provided a detailed empirical analysis of the change in the CPU and memory utilization after auto-refactoring specific code smells in isolation as well as in combination with other smells. Obtained results highlight that the refactoring techniques adopted for code smells such as god class and god method adversely affected CPU and memory usage of the application. Refactoring {\em Long Parameters} smell resulted in improvement of CPU usage but worsened memory usage. Refactoring all other code smells improved resource usage for the same workload. We noticed that applications belonging to the same category were impacted similarly by refactoring specific smells. Also, the impacts of smells on resource consumption for Java and Python applications were quite similar; hence our results can be generalized. Combined refactoring of various code smells add up to the impact of refactoring those smells individually. Based on these observations, we suggested a set of guiding principles on selecting the correct set of code smells to be refactored for the most efficient resource utilization. We also provided a mechanism based on regression analysis to accurately predict the impact of batch refactoring code smells on CPU and memory utilization before making refactoring decisions.

\bibliographystyle{ACM-Reference-Format}
\bibliography{icsemain}
\vspace{12pt}

\end{document}